\documentclass[twocolumn,showpacs,floatfix]{revtex4}
 \usepackage{graphicx}
 \usepackage{amsfonts,amssymb}
\usepackage{bm}
\begin{document}
 
\title{Stress and large-scale spatial structures in dense, driven granular flows}
\author{Allison Ferguson, Bulbul Chakraborty}
\affiliation{Martin Fisher School of Physics, Brandeis University,
Mailstop 057, Waltham, MA 02454-9110, USA}
\date{\today}

\begin{abstract}
We study the appearance of large-scale dynamical heterogeneities in a simplified model of a driven, dissipative granular system.
Simulations of steady-state gravity-driven flows of inelastically colliding hard disks show the formation of large-scale 
linear structures of particles with a high collision frequency. 
These chains can be shown to carry much of the collisional stress in the system due to a 
dynamical correlation that develops between the momentum transfer and time between collisions in these 
"frequently-colliding" particles. The lifetime of these dynamical stress heterogeneities is seen to grow as the flow 
velocity decreases towards jamming, leading to slowly decaying stress correlations reminiscent of the slow 
dynamics observed in supercooled liquids.
\end{abstract}


\maketitle

\section{Introduction}
Granular materials are ubiquitous in nature and play an important role in many industrial applications, yet the appropriate coarse-grained theories 
for the description of these materials remain elusive due to the rich phenomenology they display in response to external perturbation~\cite{jaeger96, kadanoff99}.  
A common example of the unusual nature of granular systems is the phenomenon of force chains, highly-stressed extended linear clusters first 
seen in experiments on static granular piles.~\cite{liu95}. These striking inhomogeneities are thought to be responsible for a host of observed 
nonlinear effects in static systems including interesting features in distributions of contact forces between particles~\cite{mueth98}.
However, the reason for the existence of the force chains has not been determined; one possible explanation is that correlations develop while the 
material is flowing which then get "frozen in" when the system comes to rest.

Spatial structures have been directly observed in experiments on flowing granular materials~\cite{bonamy02}. More recently, 
measurements of velocity correlations in the surface layer of particles flowing down an inclined plane have yielded a length scale 
which appears to grow as the flow is arrested~\cite{pouliquen04}.  While detection of such dynamical heterogeneities in numerical studies 
has been more difficult, there is some evidence for their existence in simulations of gravity-driven flow through a hopper~\cite{ferguson04} 
and in images of the contact force network in simulations of chute flow~\cite{silbert05}. The presence of such structures will
play a crucial role in determining the dynamics of the flowing system, especially as the transition to a static system is approached. The
development of spatial and temporal correlations near the transition to the static state must be investigated carefully in order to understand
the nature of the transition from the flowing liquid state.

Dynamical heterogeneities have been observed in other systems which exhibit a similar slow dynamics as the transition from fluid to solid 
is approached, such as colloids near the glass transition. Fast-moving particles were observed to be spatially correlated with a characteristic cluster 
size that increased as the glass transition was approached~\cite{weeks00}. In supercooled liquids, spatial inhomogeneities can be identified via a 
time-dependent four-point (two-time, two-space) correlation function~\cite{glotzer00}.  The interesting commonalities between these disparate systems
have been explored in the context of the jamming phase diagram proposed in Refs.~\cite{liu98, ohern01}. It should be noted that in these studies the focus has been
primarily on the search for a common static signature (i.e. the distribution of contact forces $P(f)$). Thus the proper framework to describe the dynamics
of systems which have a jamming transition remains an open question. If the presence of dynamical heterogeneities governs the dynamics of 
these far-from-equilibrium systems near the jamming transition, then while the microscopic process by which these structures form may vary between 
different types of systems, some unified dynamical description may be applicable once the structures have formed.  The "soft glassy rheology" (SGR) 
model~\cite{cates04,monthus96} describes soft, nonergodic systems such as foams and dense emulsions and has successfully predicted many 
of the experimentally observed results. Application of this type of "trap-like dynamics" to granular systems has also had some promising initial 
results~\cite{bouchaud02}.

In this paper we explore the connections between dynamical heterogeneities and slow dynamics in a simplified model which allows us to focus on 
the essential effects of driving and dissipation in a flowing granular system, while still reproducing observable results from related experimental 
systems~\cite{longhi02}. Simulations have been performed of a two-dimensional gravity driven system of frictionless, bidisperse hard disks 
in a hopper geometry.  The disks undergo instantaneous, inelastic binary collisions and propagate under gravity in between collision
events.  Despite the simplified dynamics of the simulation, several interesting collective effects are seen to emerge.  As in previous 
studies~\cite{ferguson04}, extended linear structures defined by inhomogeneities in the spatial distribution of the collision frequency are 
visible, and their presence has a measurable effect on distributions of impulse and time intervals between collisions for a given particle.  
In this study we observe that the particles which make up the linear structures display a dynamic correlation between their impulse and 
collision time which is reflected in measurements of the collisional stress tensor; particles identified as part of the structures appear to experience
higher than average collisional stress compared with the remainder of the system.  Using this measure of collisional stress we can probe 
further into the nature of these structures and their relevance to the dynamics of the system. 

\section{Simulations}
The grain dynamics used in the simulations are similar to Ref.~\cite{denniston99}. 
Specifically, (i) at each interparticle collision, momentum is conserved but the collisions are inelastic
so the relative normal velocity is reduced by the coefficient of restitution $\mu$; (ii) to allow the side walls 
to absorb some vertical momentum we impose the condition that collisions with the walls are inelastic 
with a coefficient of restitution $\mu_{wall}$ in the direction tangential to the wall and (iii) since we 
wish to observe the system over many events, particles exiting the system at the bottom must be 
replaced at the top to create uniform, sustained flow.  Note that collisions are instantaneous and in 
between collisions the particles are driven by gravity. To avoid the phenomena of inelastic collapse the coefficients of 
restitution $\mu$ and $\mu_{wall}$ are velocity dependent; if the relative normal velocity between particles or between a given particle and the wall is 
less than some cutoff $v_{cut}$ then the collision is presumed to be elastic~\cite{bennaim99}.  The flow velocity is controlled by adjusting 
the width of the hopper opening. We also introduce a probability of reflection $p$ at the bottom which reduces the time needed to
reach steady-state flow. Typically, our simulations were done on bidisperse systems 
(diameter ratio 1:1.2) of 1000 disks, with $\mu$ = 0.8, $\mu_{wall}$ = 0.5, $v_{cut}$ = $1\times10^{-3}$ and $p$ = 0.5. 
The simulation was run for a total time of $1\times10^3$ in simulation time units (smaller particle diameter $d_s$ and 
gravitational constant $g$ are both set to 1) with the initial time interval of $5\times10^2$ discarded 
before recording data to ensure the system has reached steady state.  During the total time interval of 
$500$ over which we are evaluating the data, a given particle will pass through the hopper 5-10 
times depending on the flow velocity.

As the width of the hopper opening is decreased the average flow velocity decreases, and at some minimum opening sustained flow is no longer 
observed.  Close to this width the mass flow exiting the hopper appears to become very intermittent, with large outflow of mass occuring
on short timescales followed by long time intervals where few particles exit the system.  Intermittent dynamics has also been observed 
in simulations of flow down an inclined plane where the angle of incline is approximately equal to the angle of repose of the grains~\cite{silbert05}.
Additionally, some preliminary measurements of spatial velocity correlations in our hopper simulations parallel to the flow direction indicate 
long range correlations on length scales comparable to the system size for slow flow velocities~\cite{tithi}.  Spatial velocity
correlations with long length scales are also seen on the surface of incline plane flow~\cite{pouliquen04}.

\section{Collision Frequency and Large-Scale Structures} 

\begin{figure}[h!]
\includegraphics[width = 0.95\columnwidth]{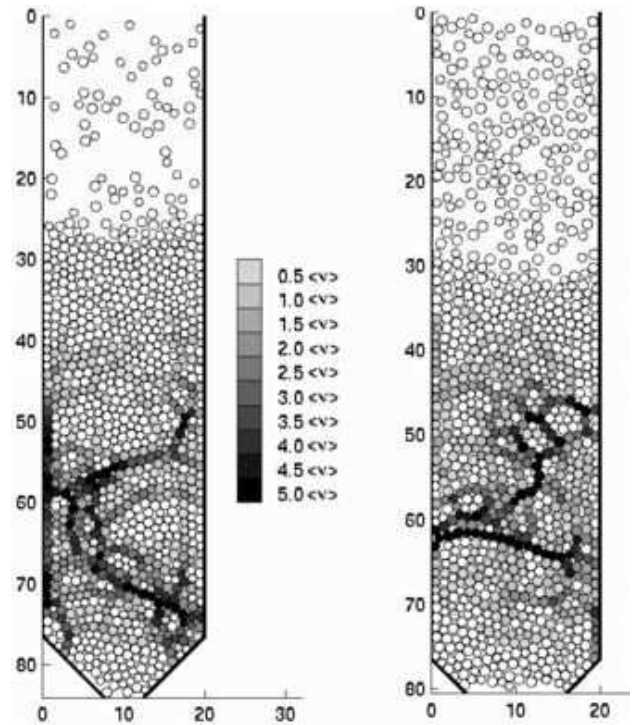}
\caption{\label{fcplot} Images of the system with particles coloured according to the collision frequency in the last
time interval $\Delta t = 0.02$ for $v_f = 0.86$ (left) and $v_f = 2.2$ (right). $\langle \nu \rangle$ is defined as the mean collision frequency (average taken
over all particles) for that time interval.}
\end{figure}

In studies of two-dimensional freely-cooling gases \cite{mcnamara94} some evidence of dynamical 
large-scale structures forming in the system was first observed by identifying all particles experiencing 
inelastic collapse as the assembly of grains cooled.  These particles formed linear structures within dense 
clusters and experienced collision frequencies significantly higher than that of the other particles in the 
system. Due to the effect of driving in our flowing system, we do not observe inelastic collapse.  
However, we can ask an analogous question: {\it ``How many collisions does a given grain undergo in 
a given time interval $\Delta t$?''} The collision frequency for a particular particle $i$ in $\Delta t$ is given by 
$\nu_i = N_i/\Delta t$ where $N_i$ is the number of collisions experienced by particle $i$ in $\Delta t$. We can construct 
images of our simulated system at regular intervals and colour individual disks according to the value of the collision frequency for that disk.  As 
can be seen in Fig.~\ref{fcplot} particles with high collision frequencies
form linear chains reminiscent of the structures observed in 2D freely-cooling gases and the transient
solid chains postulated by the hydrodynamic model of Ref.~\cite{mills99}.  These chains often appear to terminate
at the boundary of the system, forming an arch-like structure in the centre of the hopper.  From the pictures it is also evident that
collision direction is typically oriented along the chain direction (i.e. particles within a chain collide with other particles in the chain and
not with neighbouring particles which are not part of the structure).  Thus the velocities of frequently-colliding particles are 
highly correlated in direction, and we can evaluate the stability of this structure with respect to perturbations arising from collisions with 
other (uncorrelated) particles. Consider an initially isotropically distributed dense assembly of completely correlated disks and perturb one disk by 
giving it a velocity ${\bf v}$.  The collision rule for two inelastic hard disks $i, j$ is ~\cite{denniston99}:

\begin{equation}
\left[ \begin{array}{c} \bf{v'_i} \\ \bf{v'_j} \end{array} \right] = \left[ \begin{array}{c} \bf{v_i} \\ \bf{v_j} \end{array} \right] 
+ \frac{1 + \mu}{2} \left[ \begin{array}{c} \bf{-v_i \cdot  q} + \bf{v_j \cdot q} \\ \bf{v_i \cdot  q} - \bf{v_j \cdot q} \end{array} \right] \bf{q}
\label{collrule}
\end{equation}

where ${\bf q}$ is a unit vector along the centre of mass line of the colliding disks and $\bf{v'}$ denotes the velocity of a given particle after the collision.  
Rewriting Eq.~\ref{collrule} in terms of the relative velocity of particle
$j$ to particle $i$ before the collision ${\bf \delta v}(t) = {\bf v_j - v_i}$ and after the collision  ${\bf \delta v}(t+\Delta t) = {\bf v'_j - v'_i}$, 
we can obtain an approximate expression for the time derivative of the relative velocity in the limit of $\Delta t \rightarrow 0$:

\begin{equation}
\frac{d \bf{\delta v}(t)}{dt} \simeq  \frac{1+\mu}{\Delta t} \left [ \bf{-\delta v}(t) \cdot \bf{q} \right ] \bf{q}
\label{vderiv}
\end{equation}

Define $q_\parallel$ and $q_\perp$ as the components of ${\bf q}$ parallel and perpendicular, respectively, to the initial perturbing velocity.
From Eq.~\ref{vderiv} it is easy to see that $\langle \frac{d \delta v_\parallel}{dt} \rangle \simeq -C_q \langle q_\parallel \rangle$
and $\langle \frac{d \delta v_\perp}{dt} \rangle \simeq -C_q \langle q_\perp \rangle$, where $C_q = \frac{1+\mu}{\Delta t} 
\left [\bf{-\delta v}(t) \cdot \bf{q} \right ]$. However, $q_\perp$ is equally likely to be positive or negative thus $\langle q_\perp \rangle \sim 0$
while $q_\parallel$ is always positive and so $\langle q_\parallel \rangle > 0$.  Substituting this result into the above expressions
for the time derivatives yields $\langle \frac{d \delta v_\parallel}{dt}\rangle  < 0$ and $\langle \frac{d \delta v_\perp}{dt}\rangle \sim 0$ 
indicating that velocity differences parallel to the initial perturbation will decay while velocity differences perpendicular ${\bf v}$ will remain. Given this, 
and the organizing influence of the inelastic wall collisions leading to velocities becoming normal to the wall, the appearance of collisional chains anchored 
to the wall becomes plausible.

\section{Stress}

\begin{figure}[h!]
\includegraphics[width = 0.95\columnwidth]{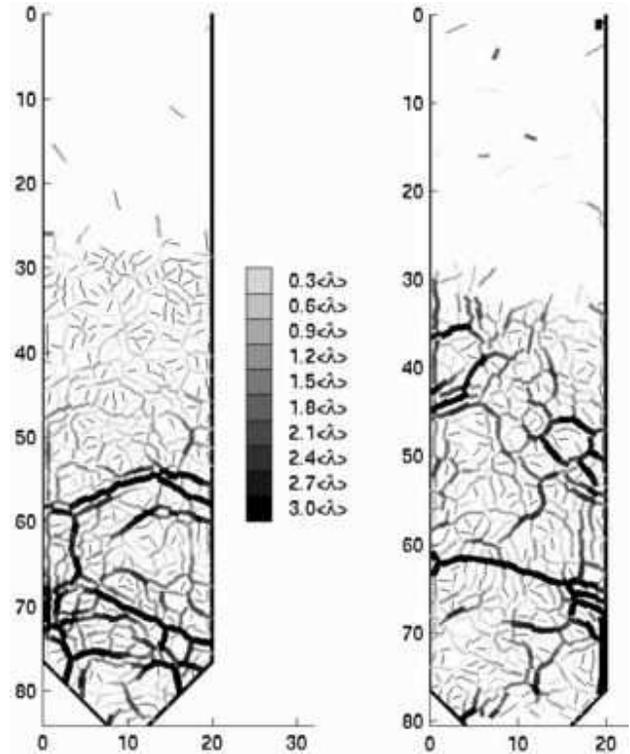}
\caption{\label{strplot}Images of the system with principal axis corresponding to the maximum eigenvalue $\lambda_m$ of the collisional 
stress tensor defined during the last time interval $\Delta t = 0.02$ for $v_f = 0.86$ (left) and $v_f = 2.2$ (right) plotted at the particle 
positions.  The principal axis line is coloured and its width scaled according to the magnitude of the maximum eigenvalue $\lambda_m$
(the thicker the line the higher the value of $\lambda_m$). Thin black lines indicate the principal axis of particles experiencing
less than the minimum stress plotted ($0.3 \langle \lambda \rangle$ for these pictures).
Note that these pictures were constructed for the same time interval as the images in Fig.~\ref{fcplot}. $\langle \lambda \rangle$ is 
defined as the mean $\lambda_m$. (average taken over all particles) for that time interval.}
\end{figure}

The unusual rheology of granular matter in response to external stress, as well as the phenomena of force chains prompts us to investigate
stress in this simple model of flowing granular matter. Specifically, are the frequently-colliding chains described in the previous section
the dynamical analog of force chains? We can explore this possible connection by measuring the stress in our system and determining if a correspondence
exists between the collisional chains and the spatial distribution of stress. Additionally, we can potentially use the measured stress to calculate relevant time
and length scales associated with the structures, as well as making a clearer connection between our model and other systems with soft, glassy rheology.

As described in Ref.~\cite{alam03}, macroscopic stress in discrete particle systems develops as a result of two microscopic mechanisms of 
momentum transfer: (a) transport of momentum due to the fluctuations of the individual particle velocities (the kinetic stress $\sigma_k$) and 
(b) transport of momentum at interparticle collisions (the collisional stress $\sigma_c$).  The total stress tensor is then the sum of the kinetic and 
collisional stresses.  However, for the densities we observe, we find as in Ref.~\cite{alam03} that the magnitude of the kinetic stress contribution is 
several orders of magnitude lower than the collisional stress.  Thus we will consider only the collisional stress contribution in the analysis to follow.
For a given particle $i$, the $\mu\nu$ component of the collisional part of the stress tensor at time $t$ is given by:

\begin{equation}
\sigma_{c,i}^{\mu\nu}(t) = \frac{1}{\Delta t} \sum_{\alpha} ({\bf I_{\alpha}}\cdot {\bf r_{i\alpha}})\hat{r}_{i\alpha}^{\mu}\hat{r}_{i\alpha}^{\nu}
\label{collstress}
\end{equation}

where $\mu, \nu$ are Cartesian coordinates.  The sum is taken over all collisions $\alpha$ experienced by particle $i$ during 
time $t - \Delta t \rightarrow t$, where the interval $\Delta t$ is chosen such that $\langle \tau \rangle \ll \Delta t 
\ll 1/v_f$ where $\tau$ is the time between successive collisions for a particular particle. This constraint ensures that the particle being evaluated 
experiences many collisions during $\Delta t$ but is not significantly rearranging relative to its neighbours. There is a separation of time scales (which
will be demonstrated more concretely in the analysis to follow) that ensures an interval $\Delta t$ satisfying this constraint is easy to locate.  
Note that for hard-disk collisions ${\bf I_{\alpha}}$ will always be 
along ${\bf r_{i\alpha}}$, therefore the ${\bf I_{\alpha}}\cdot {\bf r_{i\alpha}}$ term in the sum is simply 
$(d_i +d_j)I_{\alpha}/2$ where $d_i, d_j$ are the diameters of particle $i$ and the other particle involved in collision 
$\alpha$ respectively.

Similarly to the images of the system at a given time $t$ constructed for the collision frequency, we can make complementary
images of $\sigma_{c}(t)$ (Fig.~\ref{strplot}).  For every particle we calculate $\sigma_{c}(t)$ as described in 
Eq.~\ref{collstress}, and extract the maximum eigenvalue $\lambda_m$ along with the corresponding principal axis.  A
line along the direction of this principal axis is plotted at the particle position, and then coloured according to the value of 
 $\lambda_m$ relative to the average value of the maximum eigenvalue measured in that time interval $\langle \lambda \rangle$.  
 Fig.~\ref{strplot} shows some sample images constructed in this way for the same systems and times as
 in Fig.~\ref{fcplot}.  Comparison of the two figures reveals a strong correlation between the frequently colliding particles
 and the highly-stressed particles.  Additionally, the principal axis of the stress for the highly-stressed/frequently-colliding
 particles is typically aligned along the chain direction. There is also a lack of a obviously growing length scale as $v_f \rightarrow 0$,
however, timescales associated with these highly-stressed structures show a significant change with flow velocity as discussed below.

Note that the value of $\lambda_m$ will depend on the ratio $\sum_\alpha I_\alpha / \Delta t$, and thus, since the frequently colliding 
 particles experience many collisions in $\Delta t$ (i.e. $\nu_i \propto 1/\tau \gg \langle \nu \rangle$, where $\tau$ is the time between 
 successive collisions for a particular particle), then the sum over $I_\alpha$ can be large even if the average impulse
 per collision is small.  Thus $\lambda_m$ (and similarly $\sigma_{c}(t)$) will depend crucially on the ratio of impulse $I$ and collision time $\tau$,
 and any correlation between these quantities. Therefore, to gain a better understanding of the correspondence between the frequently-colliding particles and the 
 highly-stressed particles we have studied the joint distribution of impulse and collision time $P(I, \tau)$.

\section{Distributions of Impulse and Collision Time}

In previous work \cite{ferguson04} we modelled recent experiments on dense, gravity-driven
monodisperse granular flows\cite{longhi02}.  Both the experiments of  Ref.~\cite{longhi02} and 
the simulations we performed measured the distribution of impulses transferred at each collision, 
$P(I)$ as well as the distribution of time intervals between the instantaneous collisions $P(\tau)$.  In 
the experiment the measurement was made at the wall by using a transducer positioned a short 
distance upstream from the hopper opening to detect the impulse transfer as a function of time to that 
point on the wall. In the simulation we can separately analyze data from events located within the bulk, 
at the walls or at a specific point on the wall to mimic the experimental measurement. For most of our 
discussion we will focus on observations of distributions taken from events occurring in the bulk of 
the material.

Both the experiment and the simulation revealed the same effects in the measured 
distributions: (i) an increase in the number of small impulse events as the flow velocity is decreased as 
well as (ii) an increase in the number of small collision time events as the flow velocity is decreased.  In
the impulse distribution in particular, this change could be directly linked to the appearance of the 
frequently-colliding clusters pictured above. Additionally, a simple one-dimensional model was devised 
that demonstrated how the existence of linear clusters of particles which collide primarily along the 
chain direction would lead to the observed behaviour in $P(I)$ \cite{ferguson04}. The flow-rate 
invariant exponential form of $P(I)$ at large impulse as measured in both experiment and simulation 
was shown to be a reflection of the local velocity distribution, which had an exponential tail (note that 
the impulse distribution of uncorrelated particles is essentially the convolution of the individual velocity 
distributions).  Using the same 1D toy model we were able to demonstrate that the large impulse region 
of $P(I)$ is determined primarily by the form of the velocity distribution as observed.

\begin{figure}[h!]
\includegraphics[origin = c, angle = 270, width = 0.9\columnwidth]{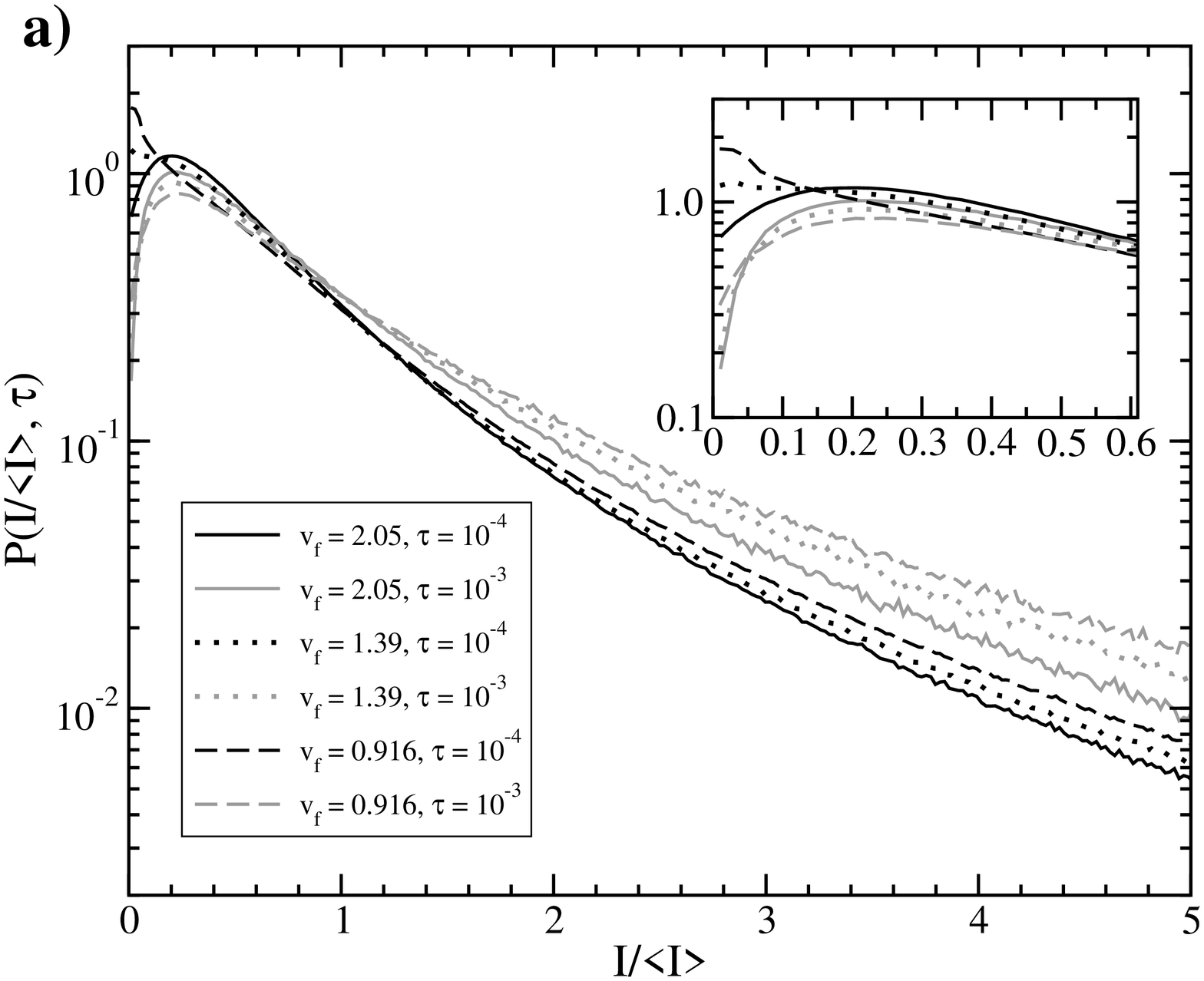}
\includegraphics[origin = c, angle = 270, width = 0.95\columnwidth]{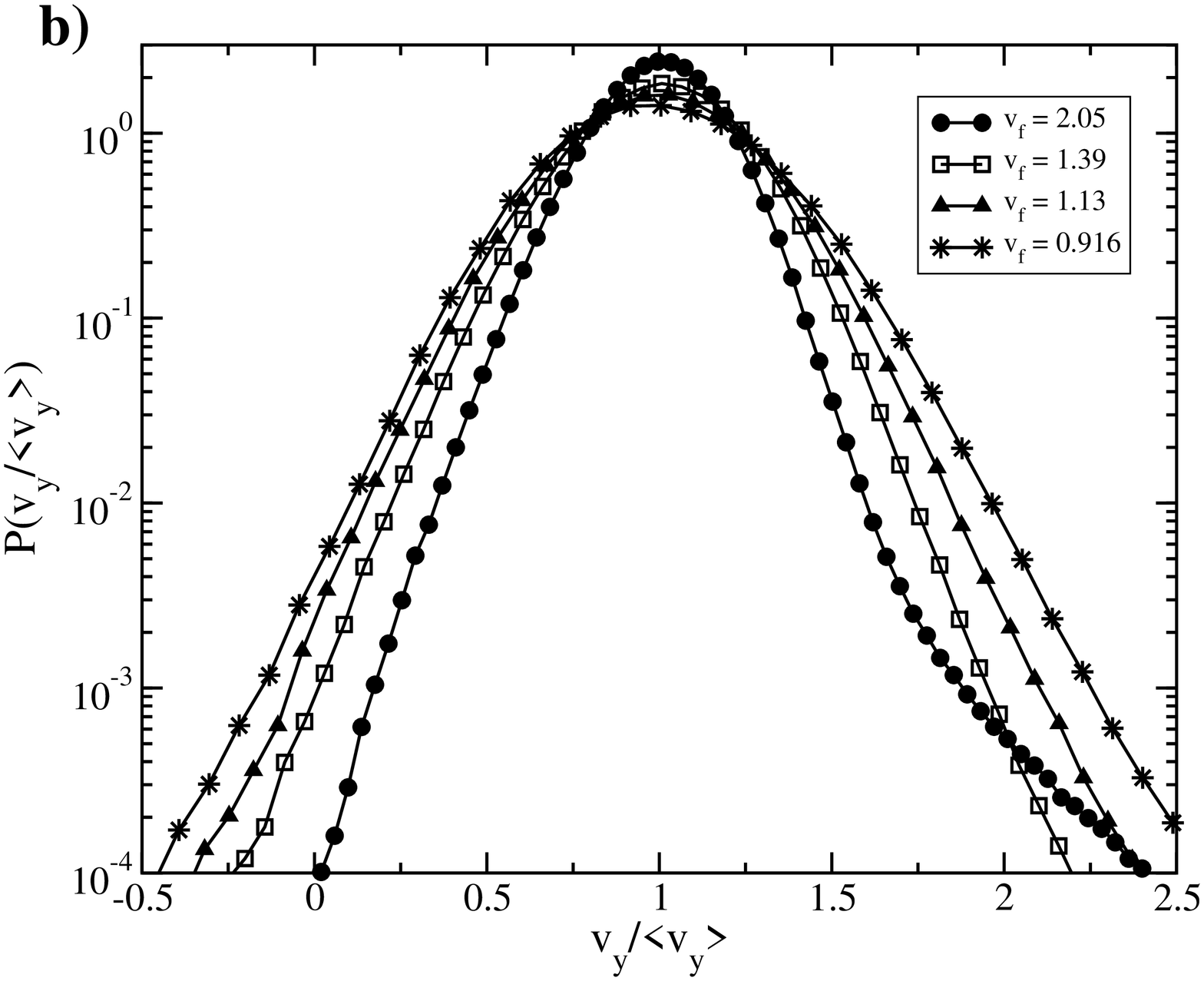}
\caption{\label{itdist_all} (a) Joint impulse-collision time distribution $P(\tilde{I}, \tau)$ at varying flow velocities. The inset shows 
the small impulse region of the distribution in more detail. Note that the two cuts of $P(\tilde{I}, \tau)$ have been
separately normalized for easier comparison. (b) Corresponding velocity distributions $P(v_y/\langle v_y \rangle)$.}
\end{figure}

\begin{figure}[h!]
\includegraphics[origin = c, angle = 270, width = 0.95\columnwidth]{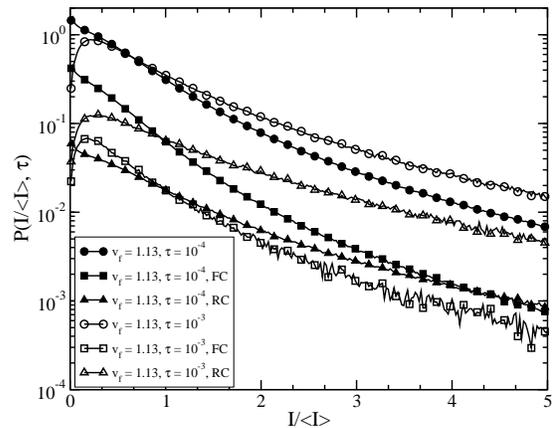}
\caption{\label{itdist_fcrc}Impulse distribution $P(\tilde{I}, \tau)$ showing the separate contribution from the frequently-colliding 
(FC) particles ($\nu_i > 5\langle \nu \rangle$) and the rarely-colliding (RC) particles ($\nu_i < 0.5\langle \nu \rangle$) at $v_f = 1.13$.}
\end{figure}

In the bidisperse system, the picture retains many of its original features with some added complexity. Defining the scaled impulse
$\tilde{I} = I/\langle I \rangle$ where $\langle I \rangle = \int I P(I) dI$ we can consider
the behaviour of the joint distribution of impulse and collision time $P(\tilde{I}, \tau)$. Shown in Fig.~\ref{itdist_all}a 
as a function of $\tilde{I}$ for two values of $\tau$, several interesting features are seen to emerge. For 
both values of $\tau$, $P(\tilde{I}, \tau)$ still shows an increase in height at impulses smaller than the average impulse as the flow 
velocity is decreased.  However, for $\tau = 10^{-3}$ the large impulse region of the distribution is no longer flow rate independent 
as in the monodisperse system, and an increase in height at impulses much larger than the average is also visible~\cite{footnote}.  As 
before, this large impulse behaviour can be linked to an accompanying shape 
change in the velocity distribution (Fig.~\ref{itdist_all}b), and is similar to the behaviour of $P(\tilde{I})$ in 
three-dimensional MD simulations of soft spheres \cite{landry04}.

If we look at $P(\tilde{I}, \tau)$ as a function of $\tilde{I}$ for a given flow velocity, it is evident that the
impulse and collision time are correlated (i.e. small impulses tend to be associated with small collision times).  To clarify the roles
played by the frequently-colliding and rarely-colliding particles in the observed shape changes in  $P(\tilde{I}, \tau)$ we can divide 
the joint impulse-collision time distribution into contributions from these two populations (shown in Fig.~\ref{itdist_fcrc}
for $v_f = 1.13$). For this purpose, based on review of the images we chose a threshold frequency of $5\langle \nu \rangle$ where 
$\langle \nu \rangle = (1/N)\sum_i^N \nu_i$ and defined all particles with collision frequency $\nu_i$ above this threshold as 
frequently-colliding.  Rarely-colliding particles are similarly defined as particles with $0 < \nu_i < 0.5\langle \nu \rangle$. It appears 
that the correlation between impulse and collision time is stronger for the frequently colliding particles, and that 
these particles are the primary contributors to $P(\tilde{I}, \tau)$ at small $\tilde{I}$ and $\tau$~\cite{footnote2}.  Additionally, as expected from the 
observations made on the large impulse region of the total $P(\tilde{I}, \tau)$, at longer collision times and large impulses the shape of the 
distribution is governed by the contribution from the rarely colliding particles (which in turn reflects the shape changes in the velocity distribution
shown in Fig.~\ref{itdist_all}b).

The impulse/collision time correlation observed in $P(\tilde{I}, \tau)$ is further evidence that velocities of frequently-colliding particles are correlated 
with small fluctuations leading to collisions with $I < \tilde{I}$.  Note that this correlation in the velocities is apparent even in the absence of any density 
inhomogeneity. The connection between the frequently colliding particles and the highly-stressed particles is not surprising in the context of the impulse/collision 
time correlation discussed above, which would give rise to values of $\lambda_m \gg  \langle \lambda \rangle$. Therefore the stress heterogeneities 
seen in Fig.~\ref{strplot} are directly related to the dynamic correlation reflected in $P(\tilde{I}, \tau)$.  Given that this correlation is seen to become 
stronger as the flow velocity decreases, it is reasonable to expect some kind of flow rate dependent behaviour in time and length scales associated with the 
 collisional stress.  As previously described, there does not appear to be any significant change in length scale of the structures with flow velocity. As a first attempt to quantify a trend in the time scale, we will calculate the autocorrelation $C_\lambda(t)$ of $\lambda_m$ and 
 determine the dependence of the time scale for the decay of  $C_\lambda(t)$ on $v_f$.
 
\section{Relaxation of Stress}

\begin{figure}[h!]
\includegraphics[origin = c, angle = 270, width = 0.95\columnwidth]{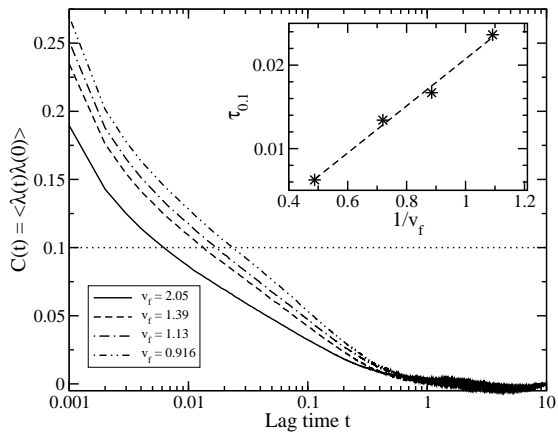}
\caption{\label{stracorr}Autocorrelation of maximum eigenvalue $\lambda_m$ of $\sigma_c(t)$.  Inset shows the time scale $\tau_{0.1}$ at which 
$\langle \lambda_m(t) \lambda_m(0) \rangle = 0.1$ plotted against $1/v_f$. }
\end{figure}

\begin{figure}[h!]
\includegraphics[origin = c, angle = 270, width = 0.95\columnwidth]{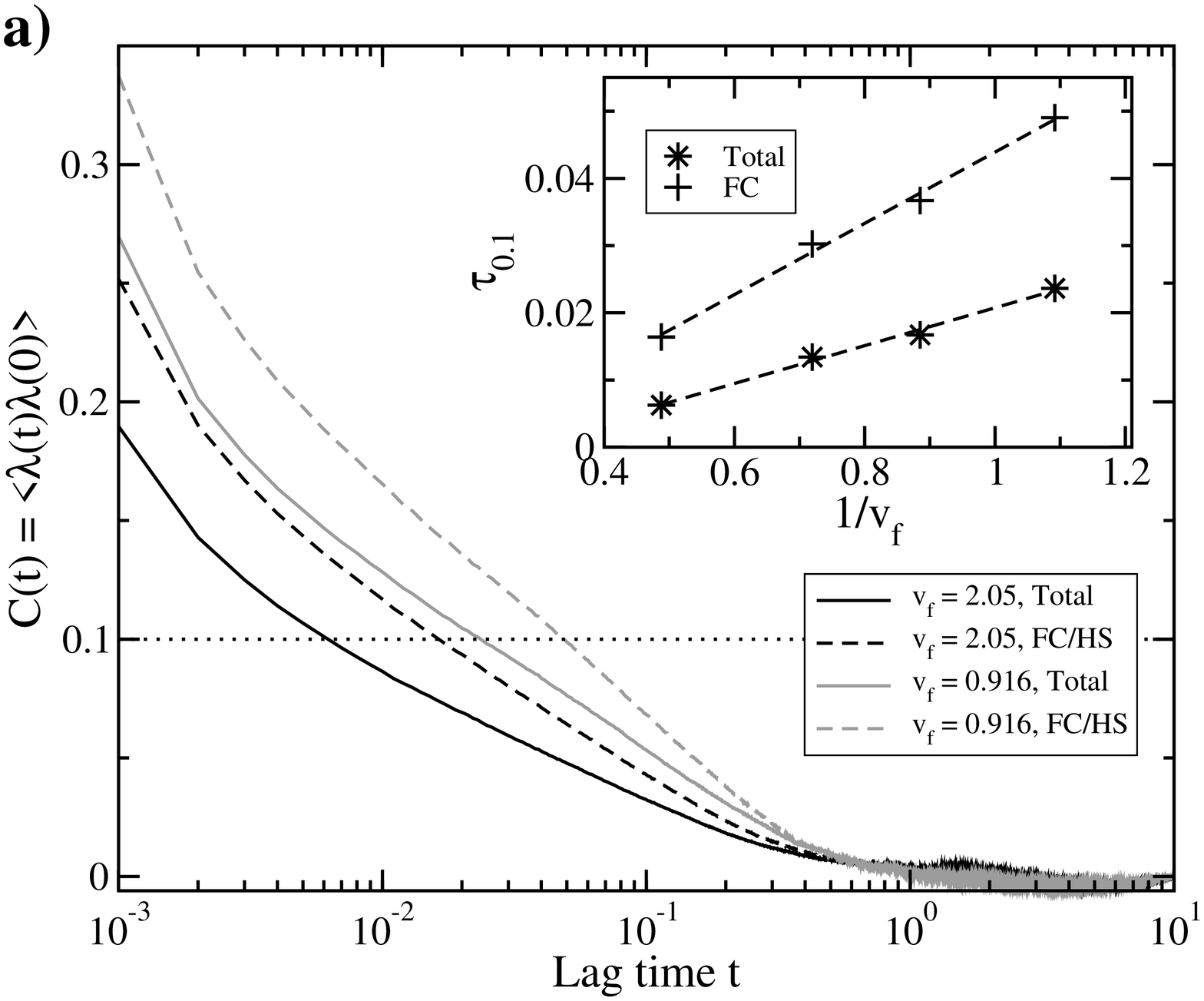}
\includegraphics[origin = c, angle = 270, width = 0.95\columnwidth]{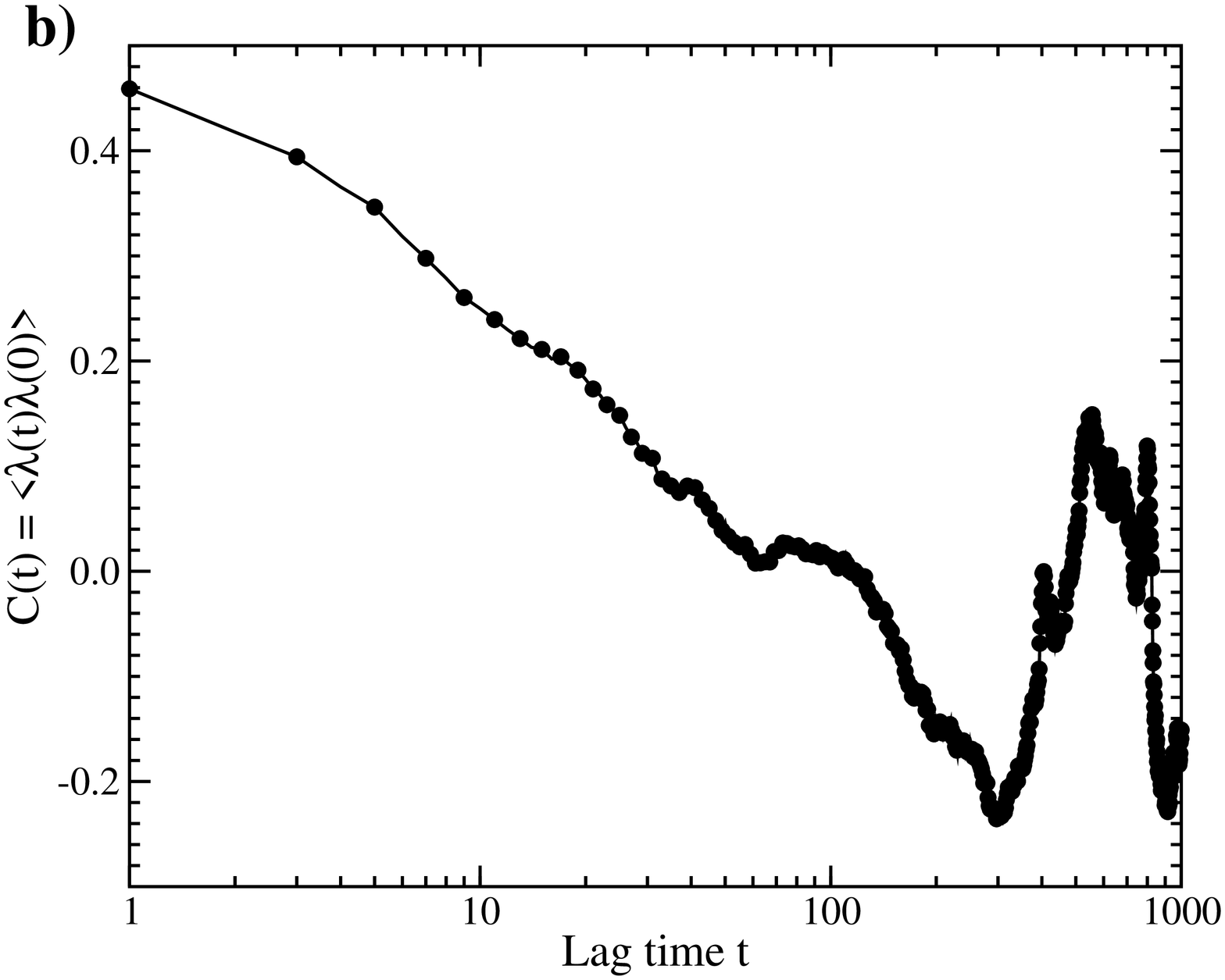}
\caption{\label{stracorr_fc} (a) Autocorrelation of maximum eigenvalue $\lambda_m$ of $\sigma_c(t)$ for both the frequently-colliding (FC) particles and the whole
system at $v_f = 2.05$ and $v_f = 0.916$.  Inset shows the time scale $\tau_{0.1}$ at which $\langle \lambda_m(t) \lambda_m(0) \rangle = 0.1$ plotted 
against $1/v_f$ for both the FC particles and the whole system. (b) Stress autocorrelation for 1D weakly driven inelastic granular gas.}
\end{figure}

The autocorrelation of the maximum eigenvalue $C_\lambda(t) = \langle \lambda_m(t) \lambda_m(0) \rangle$ of $\lambda_m$ 
is calculated as:

\begin{equation}
\langle \lambda_m(t) \lambda_m(0) \rangle = \frac{1}{N~N_{t_0}} \sum_{i, t_0} (\lambda_m(t_0) - 
\langle \lambda_m \rangle) (\lambda_m(t_0 + t) - \langle \lambda_m\rangle)
\label{acorreq}
\end{equation}

where $\langle \lambda_m \rangle$ is the time-averaged value of the maximum eigenvalue and $N_{t_0}$ is the number of
time origins $t_0$. $C_\lambda(t)$ (shown in Fig.~\ref{stracorr} for varying flow velocities) appears to have three decay regimes: 
(i) a short power-law regime at small time $t \sim \langle \tau \rangle$, (ii) a longer slow (possibly logarithmic) decay at intermediate time 
$\langle \tau \rangle \ll t \ll 1/v_f$ and (iii) a more complicated decay at long time $t \gg 1/v_f$ which involves a shallow negative dip.
This three-stage form of $C_\lambda(t)$ is reminiscent of density-density autocorrelations measured in supercooled liquids at low 
temperatures~\cite{kob02}, which show a power law regime at short time scales, a plateau at intermediate time scales
and a decay to zero at much longer time scales.  This implies that the intermediate slow decay observed in the granular system 
may be resulting from a type of "caging effect" introduced by the correlated motion of particles participating in one of the highly-stressed/
frequently-colliding chains. This type of $C_\lambda(t)$ indicates a growing separation of time scales as the flow velocity is decreased, which
is an interesting similarity between granular flows and supercooled liquids.

We can get a measure of the time scale associated with this intermediate slow decay of the stress by extracting the time  $\tau_{0.1}$ at which 
$C_\lambda(t)$ has decayed to $10\%$ of its original value (i.e $C_\lambda(\tau_{0.1}) = 0.1)$.  Plotting $\tau_{0.1}$ vs $1/v_f$ 
(see inset to Fig.~\ref{stracorr}) clearly indicates that $\tau_{0.1}$ diverges as $1/v_f$.  If we associate this time scale with the lifetime of the 
highly-stressed spatial structures visible in Fig.~\ref{strplot} then one can see that as the flow velocity decreases the structures last for increasingly 
longer times.  $C_\lambda(t)$ for the frequently-colliding particles only is shown in Fig.~\ref{stracorr_fc}a.  These curves were calculated by 
choosing only those time origins $t_0$ where a given particle had been identified as frequently-colliding according to the criteria defined above
($\nu_i > 5\langle \nu \rangle$).  Note the particle is not required to be frequently-colliding during the entire time interval over which  
$C_\lambda(t)$ is calculated, thus we expect the long time behaviour of these curves to be similar to that of the entire system.  The same 
three-regime behaviour is evident, but the time scale of the decay $\tau_{0.1}^{FC}$ is longer ($\tau_{0.1}^{FC} = 2 \tau_{0.1}$).  However, the 
scaling of the decay time with the flow velocity is the same (see inset to Fig.~\ref{stracorr_fc}a).

The role of the frequently-colliding/highly-stressed
particles in determining the shape of the autocorrelation can be clarified by considering a related system; a one-dimensional assembly of particles 
which experience instantaneous, binary, inelastic collisions.  To establish a steady-state, a fraction of the particles are weakly driven in between collision
events.  This 1D system represents an idealized, isolated analog to the frequent-collision chains, and thus the stress autocorrelation measured in this system
should be representative of the stress relaxation behaviour of these structures (note that $\sigma_c(t)$ in the 1D system is a scalar quantity).  
$C_\lambda(t)$ for the 1D system is shown in Fig.~\ref{stracorr_fc}b~\cite{rajesh}.  From this plot it can be seen that the initial stress relaxation
is slow and possibly logarithmic, similar to the slow decay regime of $C_\lambda(t)$ in the 2D flowing system.  From this result it 
appears that the observed dynamical heterogeneities do dominate the stress relaxation in the flowing granular system at intermediate time scales 
$\langle \tau \rangle \ll t \ll 1/v_f$.

\section{Conclusions}
The picture which is emerging from these simulations is that large-scale highly-stressed structures analogous to force chains in static systems can 
form even in a simplified model of inelastic hard disks. These structures experience a slow relaxation from collisional stress at intermediate time scales 
in a manner analogous to temporal relaxations observed in glassy systems at low temperatures.  Additionally, the time scale extracted from measurements 
of the stress relaxation curve $C_\lambda(t)$ is seen to diverge as $1/v_f$. This time scale can be associated with the average lifetime of a 
highly-stressed/frequently-colliding chain in the system and thus the chains appear to become infinitely long-lived as $v_f \rightarrow 0$.

It is worth noting that a similar search for a relevant length scale associated with these chains has been more difficult.  Simple two-point spatial correlations
of the collisional stress have not yielded any meaningful results and thus a higher-order correlation function such as the four-point correlations measured in
supercooled liquids~\cite{glotzer00} measured may be necessary.  Initial calculations of spatial velocity correlations in our flowing system have indicated 
the existence of a length scale which is increasing as the flow velocity is decreased~\cite{tithi}. Preliminary measurements show that this length scale is 
of the order of the system size at the slowest flow velocities measured thus far.  Therefore another important consideration is finite-size effects; 
while we have not explored this issue in this preliminary work, some finite-size scaling may be useful in accurately determining any measurable 
length scale~\cite{berthier03}.

As discussed above, the trap model picture may provide a framework for describing the dynamics of our simple granular flow system as well as
other supercooled liquids.  To explore this possibility the relevant quantity corresponding to the distribution of trapping times must be determined 
and measured, and the predictions of this type of coarse-grained dynamics quantitatively examined.

\acknowledgments
We thank R. Ravindran, S. Tewari, N. Menon, T. Witten and P. Sollich for many helpful discussions. AF and BC acknowledge support from NSF through grant
No. DMR 0207106, and AF acknowledges support from the Natural Sciences and Engineering Research Council, Canada.


\begin{thebibliography}{0}
\bibitem{footnote} This change in shape at large impulses is also visible in the impulse distribution $P(\tilde{I}) = \int P(\tilde{I}, \tau) d\tau$.
\end{thebibliography}
\end{document}